\documentclass[aps,twocolumn, floats,preprintnumbers,showpacs,nofootinbib,prl]{revtex4}
\usepackage{graphicx}
\usepackage{url}
\usepackage{color}
\usepackage{enumerate}
\usepackage{amsmath}
\usepackage{amsfonts}
\usepackage{amssymb}

\newcommand{\bi}{\begin{itemize}}
\newcommand{\ei}{\end{itemize}}
\newcommand{\be}{\begin{equation}}
\newcommand{\ee}{\end{equation}}
\newcommand{\bea}{\begin{eqnarray}}
\newcommand{\eea}{\end{eqnarray}}

\def\tt{t\bar t}
\def\bb{b\bar b}
\def\tautau{\tau^+\tau^-}
\def\bbaa{b\bar b \gamma\gamma}
\def\bbll{b\bar b \tau^+\tau^-}

\def\lhhh{\lambda^{hhh}}
\def\lhhH{\lambda^{hhH}}

\date{\today} 

\begin{document}
\title{New physics in resonant production of Higgs boson pairs}
\author{Vernon Barger$^{a}$, Lisa L.~Everett$^{a}$, C.~B.~Jackson$^{b}$, Andrea~Peterson$^{a}$, Gabe Shaughnessy$^{a}$\\
\vspace{2mm}
${}^{a}$Department of Physics, University of Wisconsin, Madison, WI 53706, USA
${}^{b}$Department of Physics, University of Texas at Arlington, Arlington, TX 76019, USA
}

\begin{abstract}
We advocate a search for an extended scalar sector at the LHC via $hh$ production, where $h$ is the 125 GeV Higgs boson.  
A resonance feature in the $hh$ invariant mass is a smoking gun of an $s$-channel heavy Higgs resonance, $H$.  With one $h$ decaying to two photons and the other decaying to $b$-quarks, the resonant signal may be discoverable above the $hh$ continuum background for $M_H<$ 1 TeV.   
The product of the scalar and top Yukawa couplings can be measured to better than $10-20\%$ accuracy, and its sign can be inferred from the $hh$ lineshape via interference effects. 

\end{abstract}
\maketitle

{\it Introduction.} The discovery of the 125 GeV Higgs boson at the LHC marks the successful completion of the long quest to validate the spontaneous breaking of the Standard Model (SM) gauge symmetry.  The discovery intensifies the search for new physics at the TeV scale using this Higgs particle as a probe. 
The Higgs branching fractions have been measured in many channels, with the best determinations in $h \to \gamma\gamma$, $h \to ZZ^* \to \ell^+\ell^-\ell^+\ell^-$, $h \to WW^* \to \ell^+\nu \ell^-\bar\nu$, and (to a lesser extent) in $h \to \bb, \tautau$~\cite {ATLSummary,CMS:2014ega}. The $h$ couplings to $\tt$ and $gg$ have been inferred from the cross sections and the total Higgs width has been bounded above the SM prediction. \cite{ATLSummary,CMS:2014ega,Barger:2012hv,Dawson:2013bba,Barger:2013ofa}.  

Although all evidence to date is consistent with the SM Higgs sector, LHC14 is quickly approaching.  The precision of Higgs coupling measurements  may reach a level at which the presence of new physics contributions is evident, and an International Linear Collider could provide still greater precision~\cite{Dawson:2013bba,Barger:2013ofa}.  However, new physics may be difficult to detect in single Higgs production, for example in the decoupling limit of an extended scalar sector.  Indeed, analyses have been done with LHC8 data~\cite{CMS:2014ipa}.  Whether or not this occurs, the measurement of the $M_{hh}$ distribution  in double Higgs production at LHC14 and its model independent interpretation could give an important new way to determine whether just the SM Higgs is the whole story.  This is the aim of this study.

Double Higgs production is potentially discoverable in several final states, of which the most promising is $hh \to b\bar b \gamma\gamma$, allowing a measurement of $\lhhh$ to $O(50\%)$ from the moderate $M_{\bbaa}$ region~\cite{Yao:2013ika,Barger:2013jfa}.  The presence of $\gamma\gamma$ allows severe rejection of background, but with an attendant high cost to the signal rate.  The branching fraction of $hh\to \bbaa$  is only 0.28\% in the SM, and thus the identification of the SM signal above the continuum background is challenging.  The $hh\to \bbll$ signal is rendered unfeasible by the reducible background of $b\bar b jj$ where both light flavored jets fake a $\tau$.  The $b\bar b W^+W^-$ channel has only a small significance in the SM~\cite{Baglio:2012np}.  

The small SM $hh$ cross section provides an opportunity to search for new physics in this channel.  In models with two (or more) Higgs doublets, the $hh$ cross-section may be enhanced significantly above the SM by way of an $s$-channel Higgs boson resonance, $H$, dramatically improving the opportunity for a new physics discovery at the LHC.   Measurements of the scalar mass and couplings can lead to a deeper understanding of the scalar sector with implications, for instance, in vacuum (meta)stability~\cite{Degrassi:2012ry}.

\begin{figure}[b!]
\centering
\includegraphics[width=0.4\textwidth]{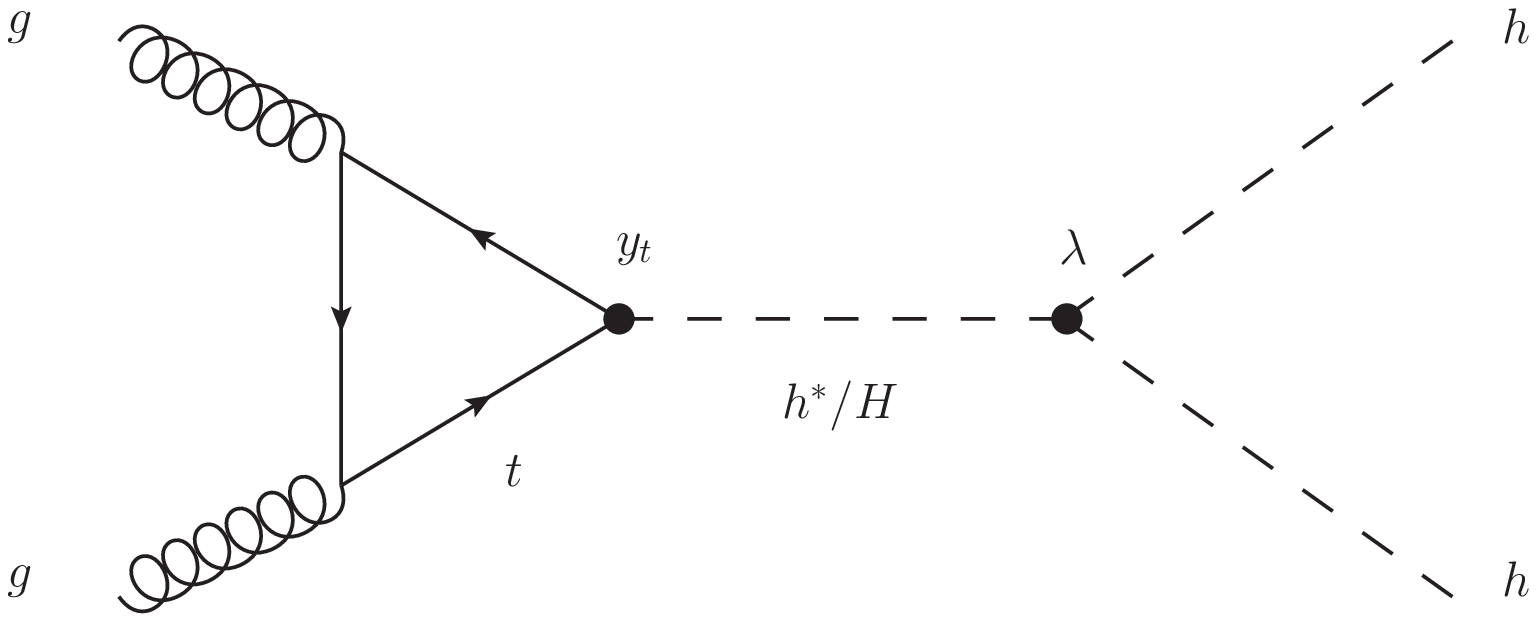}
\includegraphics[width=0.4\textwidth]{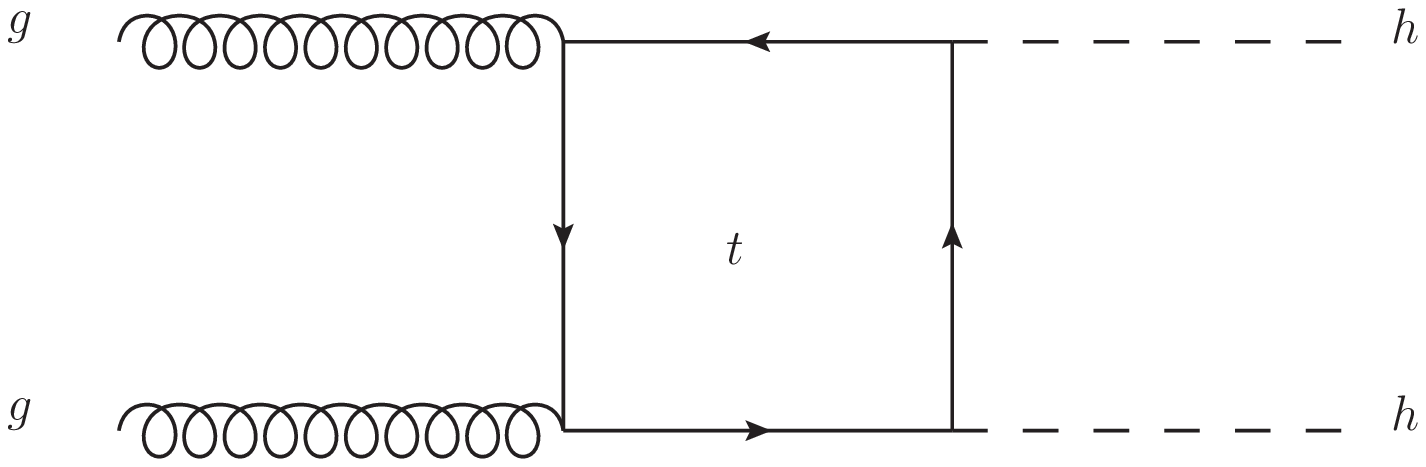}
\caption{Representative processes that contribute to Higgs boson pair production. The final state requires a $0^+$ resonance.}
\label{fig:diags}
\end{figure}
{\it Higgs Pair Production.} The Higgs pair is dominantly produced through two classes of amplitudes~\cite{Eboli:1987dy,Glover:1987nx,Dicus:1987ic}: ({\it i}) the  triangle diagram in which an $s$-channel $J^{CP}=0^+$ particle mediates the gluon-gluon transition to two Higgs bosons and ({\it ii}) the box diagram in which the annihilation of two gluons through a loop produces the Higgs boson pair, as shown in Fig.~(\ref{fig:diags}).  These amplitudes with generic internal/external Higgs bosons (and generic heavy quarks) were first computed in~\cite{Plehn:1996wb}, to which we refer readers interested in the details. 

The differential cross section at the parton level is
\bea
\frac{d\hat{\sigma}_{gg \to hh}}{d\hat{t}} &=& K \frac{G_F^2 \alpha_s^2}{256 (2\pi)^3} \\
&\times&\biggl[ \bigg|  \left( C_\triangle F_\triangle + C_\square F_\square \right) \biggr|^2 + \bigg| C_\square G_\square \biggr|^2 \biggr] \nonumber\,,
\eea
in which the $C$'s are coupling form factors, the $F$'s and $G$'s are gauge invariant form factors arising from orthogonal gluonic tensor structures, and the NLO+NNLL K-factor is  $K=2.27$ at 14 TeV~\cite{Dawson:1998py,Dittmaier:2011ti,Branco:2011iw,Shao:2013bz,deFlorian:2013uza,deFlorian:2013jea}.  Resonant production can shift the overall K-factor, as $\sigma_{NLO+NNLL}/\sigma_{LO}$ can be $\sqrt{\hat s}$ dependent. However, since the K-factor has not been calculated for the resonant process, we adopt the SM value under the reasonable assumption that any change in its value due to  the $H$ resonance is small.

\begin{figure}[h]
\begin{center}
\includegraphics[width=0.4\textwidth]{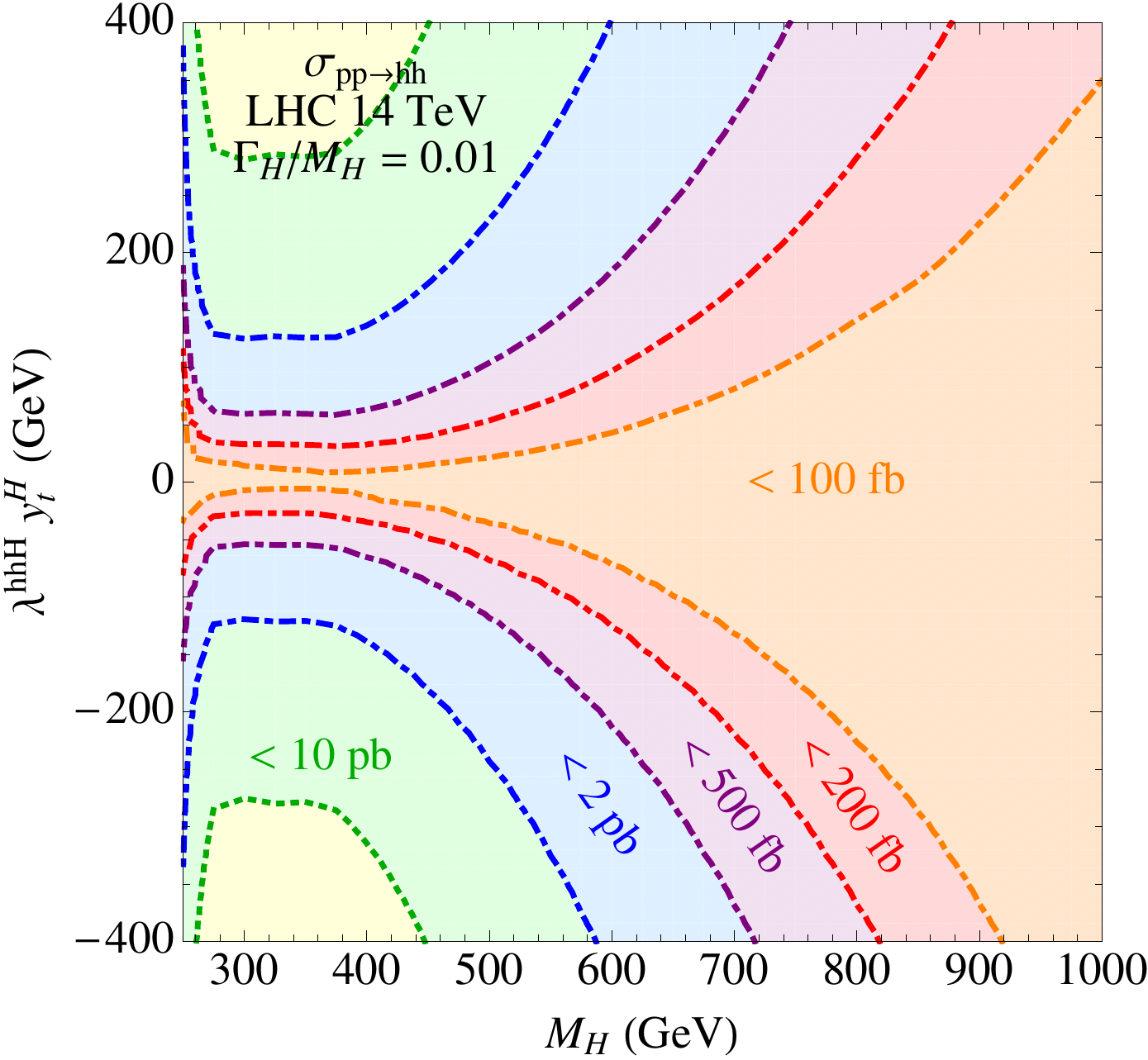}
\caption{Higgs pair production cross section at LHC14 with $\Gamma_H / M_H = 1\%$. $\sigma^{\rm SM}_{pp\to hh}= 43$ fb (with a NLO+NNLL K-factor).}
\label{fig:xs}
\end{center}
\end{figure}

The $s$-channel diagram in Fig. 1 contains the $h$ and $H$ exchanges, where $H$ is the heavy resonance that must be a $0^+$ state.  Note that the particle(s) that couple to the SM doublet $\Phi$ need not carry quantum numbers, and can couple to quarks via mixing.  
We denote the coupling of the $H$ to the $hh$ pair as
\be
{\cal L} \supset - \lambda^{hhH} H  \Phi^\dagger \Phi.
\ee
The SM tri-scalar coupling has the form $\lambda^{hhh}_{\rm SM} = {3 M_h^2/ v}$. The coupling combination $ \lhhH y_t^H$, where $y_t^H$ is the $t$-quark Yukawa coupling to $H$, is the prefactor in the $s$-channel $H$ amplitude.  The $hh$ cross section is shown in Fig.~\ref{fig:xs} for LHC14.  The enhancement from the resonance is typically over 100 fb, well above the SM value of $\sigma^{\rm SM}_{pp\to hh}= 43$ fb \cite{Shao:2013bz}.  For a sufficiently heavy $H$, the $\lhhh$ uncertainty is approximated to be the SM uncertainty~\cite{Barger:2013jfa}; there is a strong dependence of the coupling uncertainty at low $M_{hh}$, where the SM signal is strong.

{\it Measuring the $hh$ lineshape.} 
In the $\bbaa$ channel,  the dominant irreducible background is the continuum $pp\to \bbaa$ process.  We also include the additional processes listed in~\cite{Barger:2013jfa} and adopt their tagging efficiencies, mis-tag rates, threshold requirements and momentum smearing.  We simulate the resonant production of $hh\to \bbaa$ 
 and as minimal cuts, require two $b$-tags, two $\gamma$ tags and $|M_{b b} - m_h| < 20$ GeV and $|M_{\gamma\gamma} - m_h| < 10$ GeV. Additional details on the acceptance criteria are given in~\cite{Barger:2014xxx}.

We first study resonant production alone, which serves to determine the response of the detector simulation to an injection of $hh$ states at a given $\sqrt{\hat s}$, greatly simplifying the subsequent analysis.  The analytic differential distributions are Gaussian smeared  according to the resonant shape after event simulation including particle identification and isolation.  In the range of masses we consider, $M_H=200-1000$ GeV, the energy resolution of the reconstructed $\bbaa$ resonance is found to vary with increasing $M_H$ between $5-25$ GeV.  The acceptance is computed for each resonant mass, which ranges from 6-11\%.  

To determine how well the $H\to hh$ resonance signal can be measured at LHC14, we perform a $\Delta \chi^2$ fit of the $M_{hh}$ distribution to the expected SM $pp\to hh$ distribution that does not contain a resonance.  We also include the effect of increased uncertainty via background subtraction by inflating the Poisson uncertainty according to the continuum $\bbaa$ background distribution, and assume a 100\% background normalization uncertainty.

\begin{figure}[htbp]
\begin{center}
\includegraphics[width=0.4\textwidth]{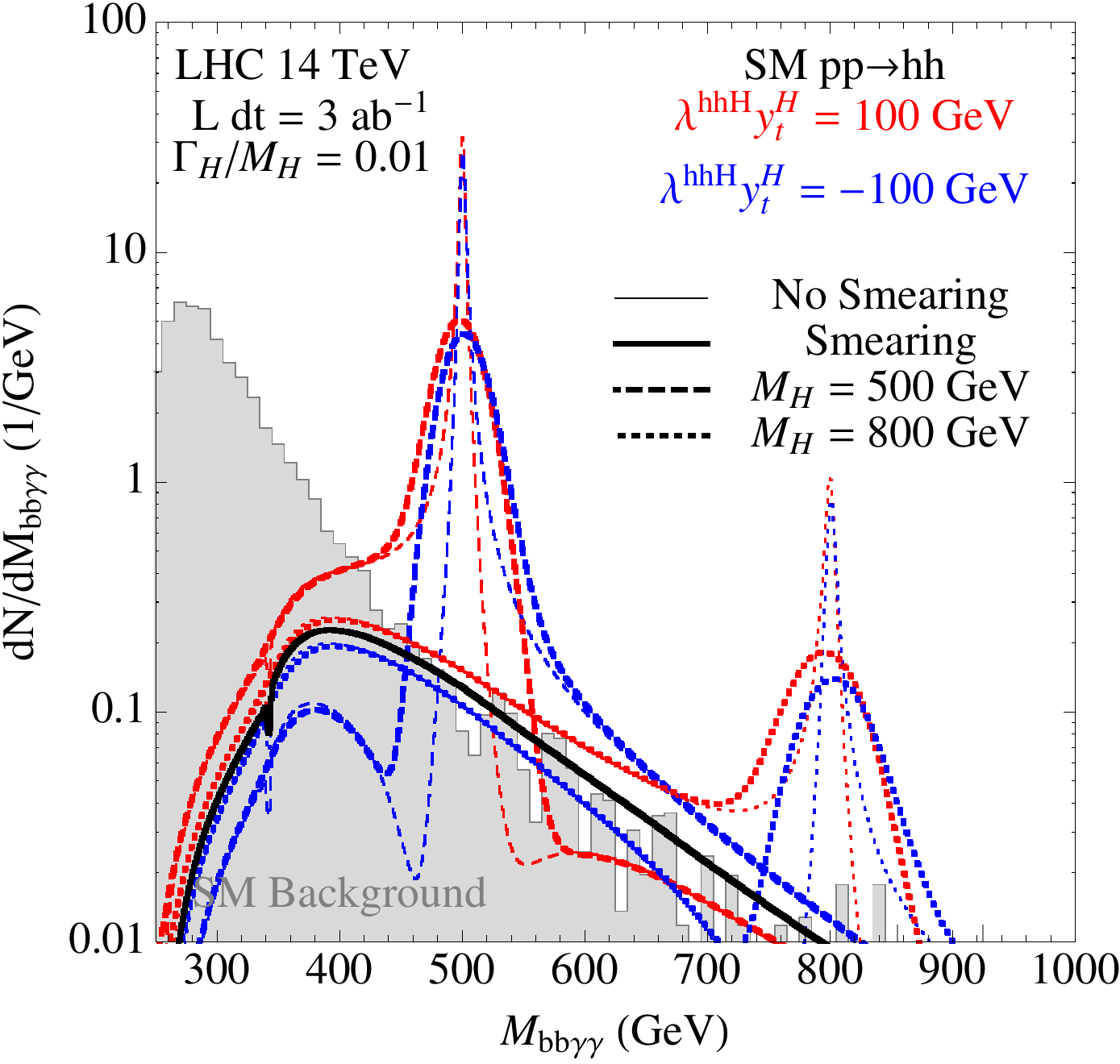}
\includegraphics[width=0.4\textwidth]{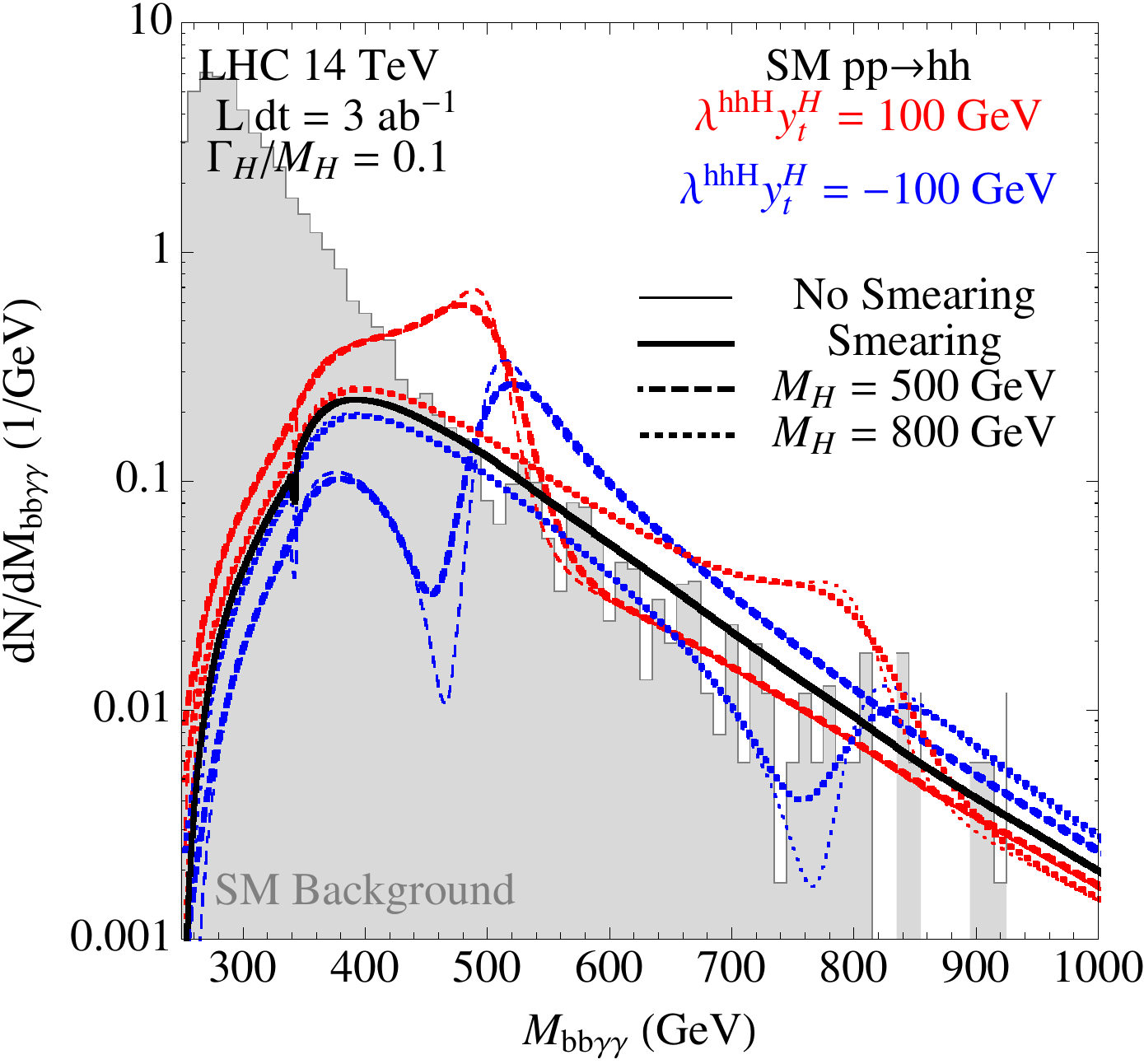}
\caption{The ${dN/ dM_{\bbaa}}$ distribution for  $M_H=500,800$ GeV  and $ \lambda^{hhH} y_t^H= \pm 100$ 
GeV, for $\Gamma_H/M_H=1\%$ (top) and $\Gamma_H/M_H=10\%$ (bottom). The SM $\bbaa$ continuum background (gray shaded) peaks at low invariant mass while the SM $hh$ continuum is denoted by the black curve.  Even with momentum smearing, the sign of $ \lambda^{hhH}y_t^H$ can be resolved.}
\label{fig:dist}
\end{center}
\end{figure}
The differential event rate is computed via
\bea
{d\sigma_{\bbaa}\over d M_{\bbaa}} = 2\,\text{BF}(h\to b\bar b)~ \text{BF}(h\to \gamma\gamma)  A(M_{\bbaa}) \times \\
\int dM^\prime G(M^\prime, M_{\bbaa}) {d\sigma_{gg\to hh}\over d M^\prime}\int_{\left(M^\prime\right)^2 / s}^{1} d\tau {d{\cal L}_{gg}\over d \tau},\nonumber\,
\eea
where $A(M_{\bbaa})$ is the signal acceptance with the resonance at $M_H=M_{\bbaa}$, the Gaussian kernel with energy resolution $\sigma$ is $G(E^\prime,E)=(\sqrt{2\pi} \sigma)^{-1}e^{-(E^\prime-E)^2\over 2 \sigma^2}$, and $ {d{\cal L}_{gg}/d \tau}$ is the parton luminosity for $gg$ collisions.

In addition to  $\lambda^{hhh}$, the  two free parameters that determine the differential distribution are $M_H$ and $ \lambda^{hhH}y_t^H$.  In Fig.~\ref{fig:dist}, we show the differential distributions for $M_H = 500\; (800)$ GeV with $\Gamma_H / M_H = 1\% \; (10\%)$ for $ \lambda^{hhH}y_t^H = \pm100$ GeV.  The total width of $H$ can vary considerably depending on $\lambda^{hhH}$ and its coupling to gauge bosons. Since the only measurable quantity is $\lambda^{hhH}y_t^H$, the production rate can be taken to be independent of the total width, and therefore we just take the two cases.

\begin{figure}[t]
\begin{center}
\includegraphics[width=0.4\textwidth]{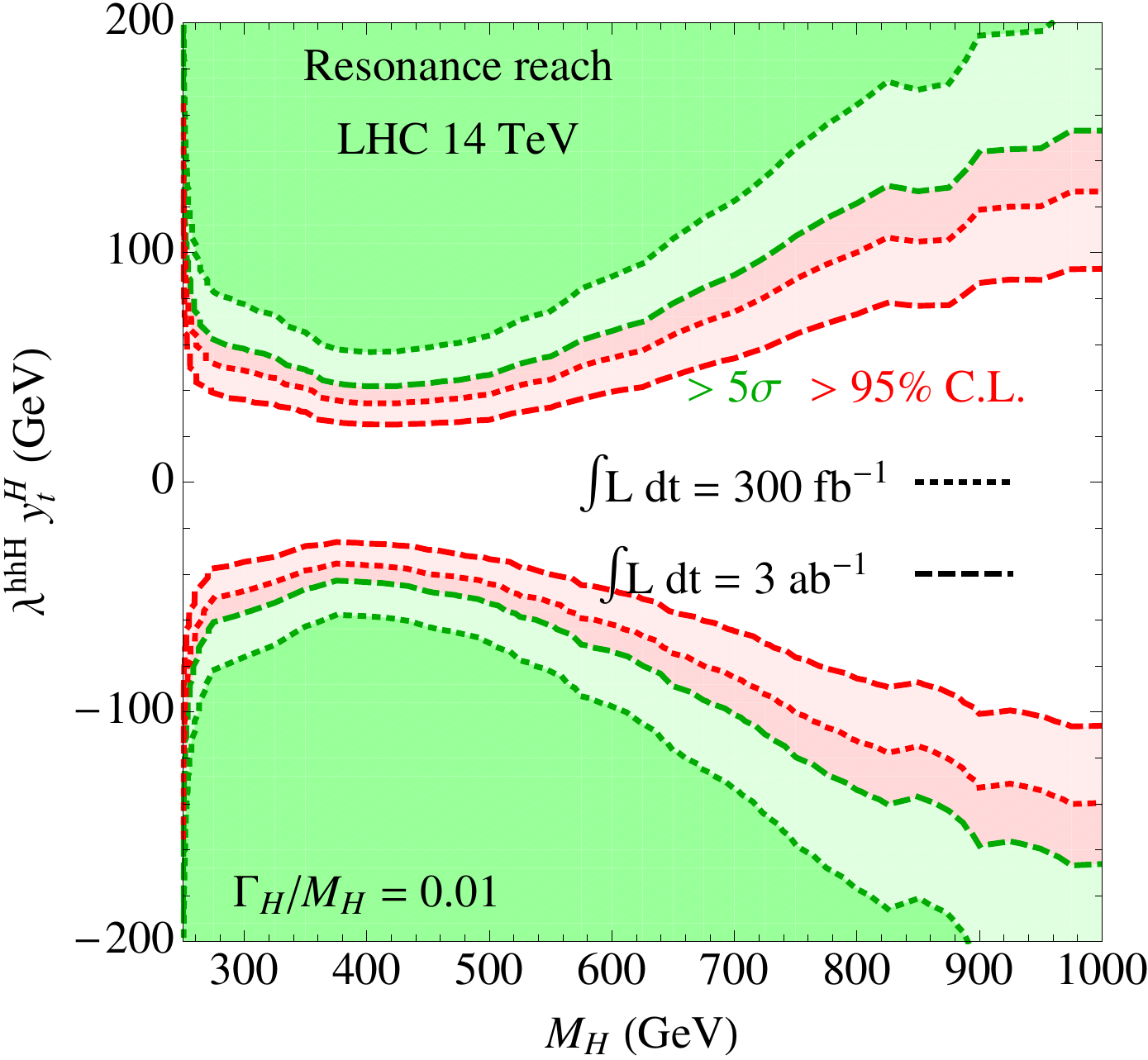}
\includegraphics[width=0.4\textwidth]{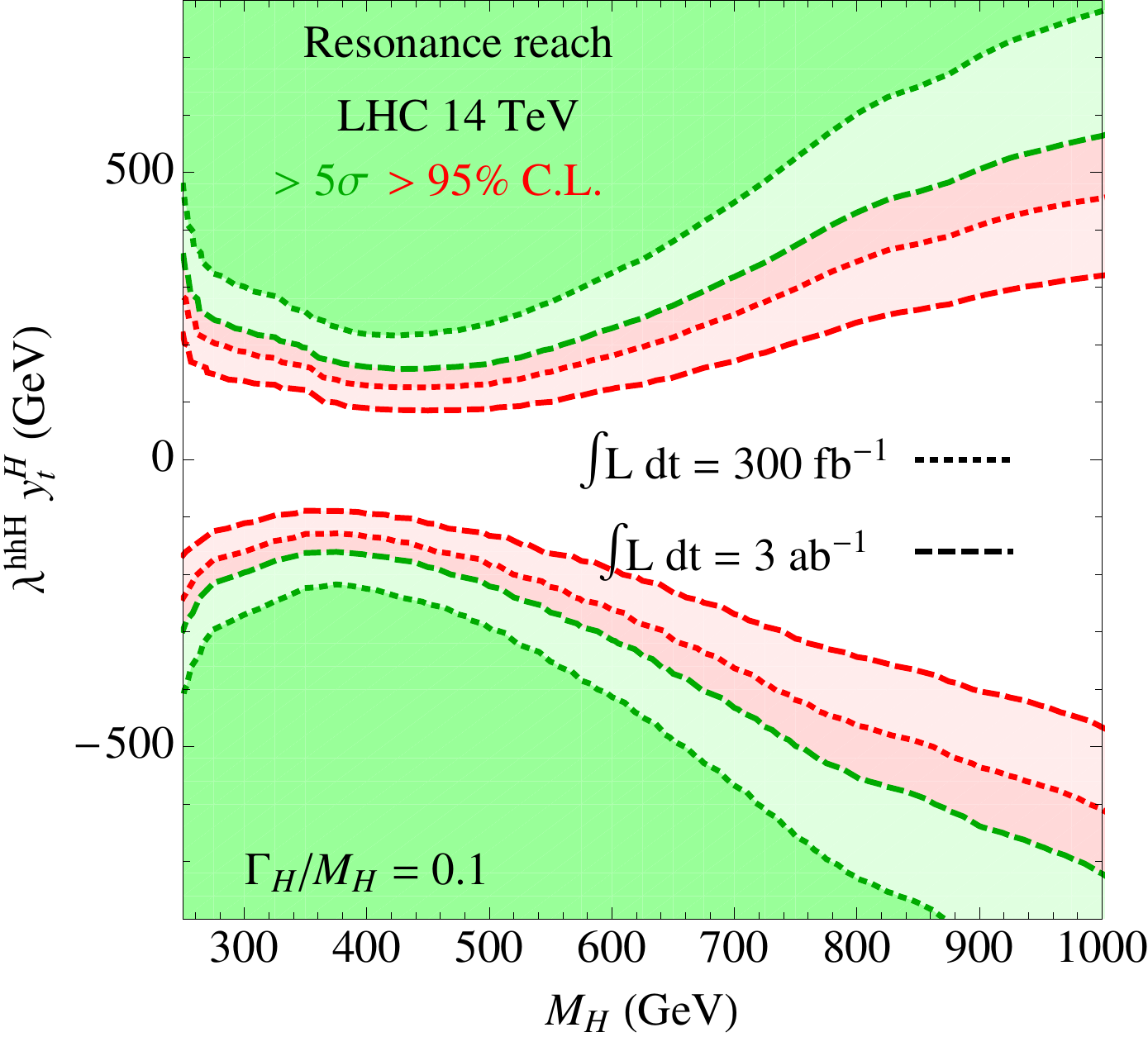}
\caption{LHC14 reach of the $H$ resonance in the $M_H$ and coupling plane with $\int {\cal L} dt=300$ fb$^{-1}$ 
for $\Gamma_H / M_H = 1\%$ (top) and $\Gamma_H / M_H = 10\%$ (bottom). The asymmetry in the contours is due to the SM continuum contribution.  }
\label{fig:reach}
\end{center}
\end{figure}
In Fig.~\ref{fig:reach}, we see that with $\int {\cal L} dt=300$ fb$^{-1}$, 
LHC14  can constrain $| \lambda^{hhH}y_t^H| \gtrsim 100$ (500) GeV over most of the mass range up to 1 TeV for a 1\% (10\%) total width.  The decrease in the reach for low $M_H$ is mainly due to the continuum $\bbaa$ background at low $M_{\bbaa}$ and (to a lesser extent) the reduced phase space near threshold.

\begin{figure}[htbp]
\begin{center}
\includegraphics[width=0.4\textwidth]{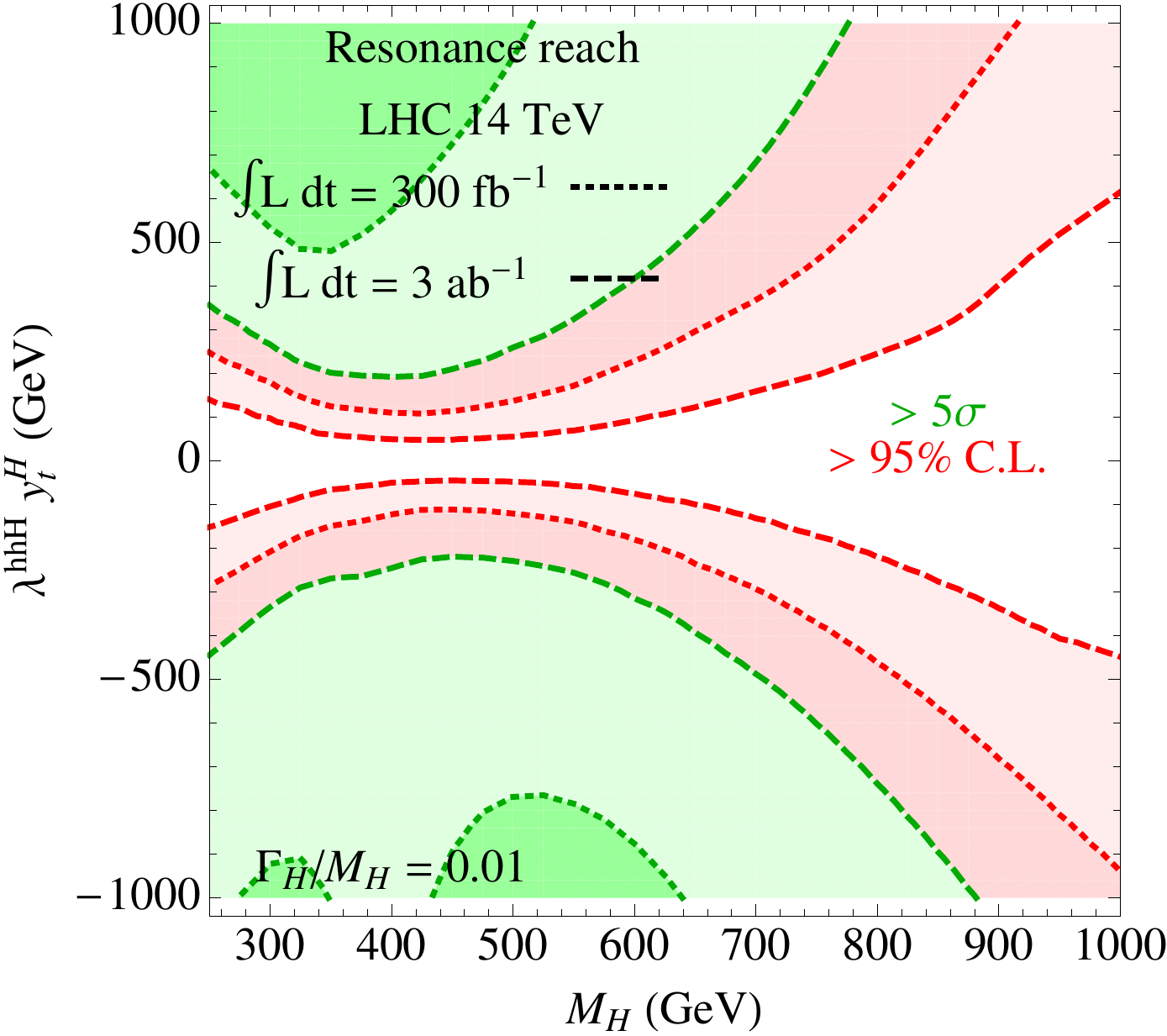}
\includegraphics[width=0.4\textwidth]{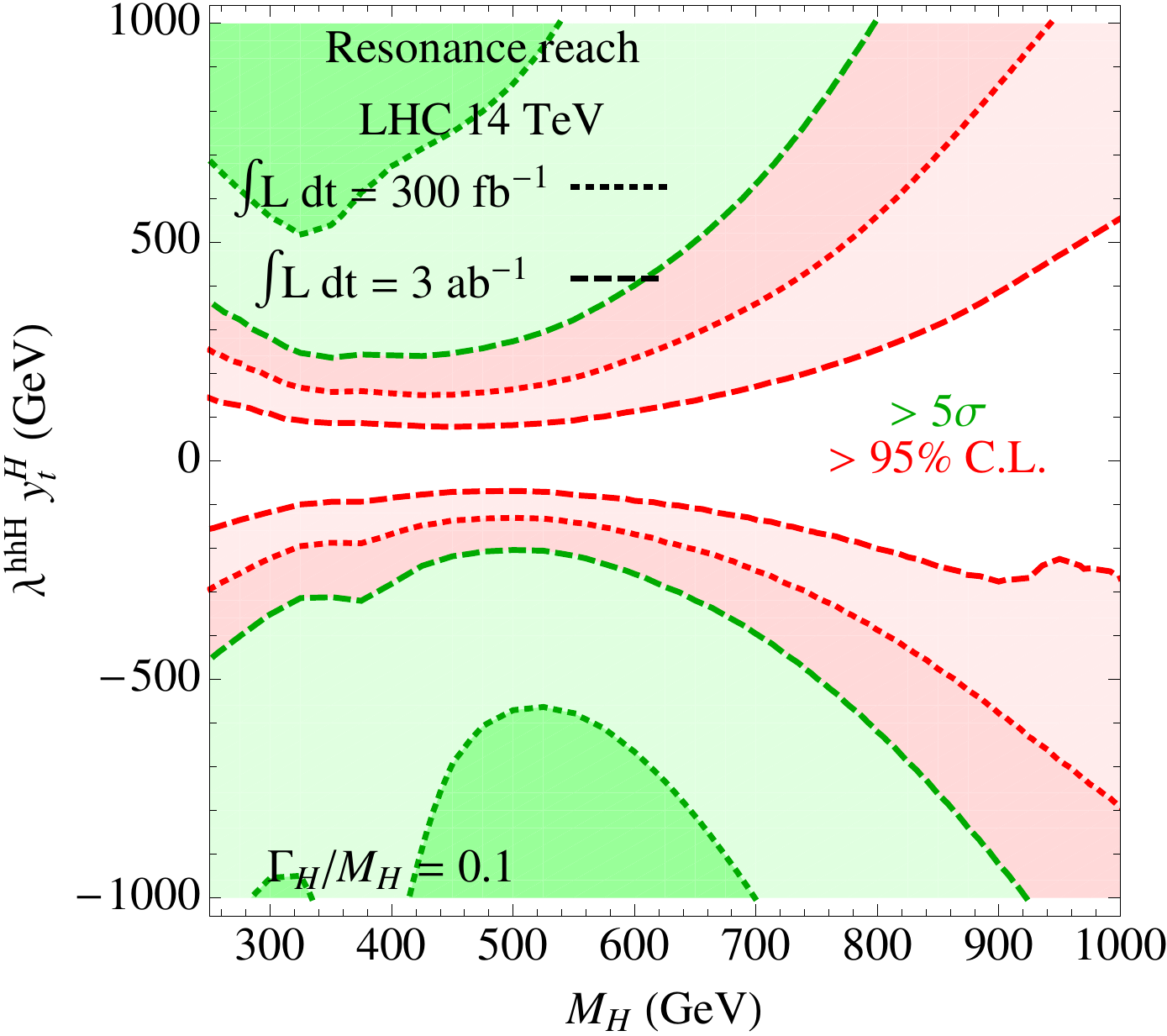}
\caption{LHC14 sensitivity to the sign of $ \lambda^{hhH}y_t^H$ from the interference between the SM and $H$ resonant diagrams for $\Gamma_H / M_H = 1\%$ (top) and $\Gamma_H / M_H = 10\%$ (bottom).} 
\label{fig:sign}
\end{center}
\end{figure}

The sign of $ \lambda^{hhH}y_t^H$ can also be determined from the interference of the resonant and continuum amplitudes. This is seen in Fig.~\ref{fig:dist} where, even with momentum smearing, the interference effect is clearly visible.  For $ \lambda^{hhH}y_t^H>0$, an excess of events occurs below resonance and a deficit above resonance (and vice-versa for $ \lambda^{hhH}y_t^H<0$). In Fig.~\ref{fig:sign}, we show regions over which the sign of $ \lambda^{hhH}y_t^H$ can be distinguished via the $hh$ lineshape.  

\begin{figure}[htbp]
\begin{center}
\includegraphics[width=0.4\textwidth]{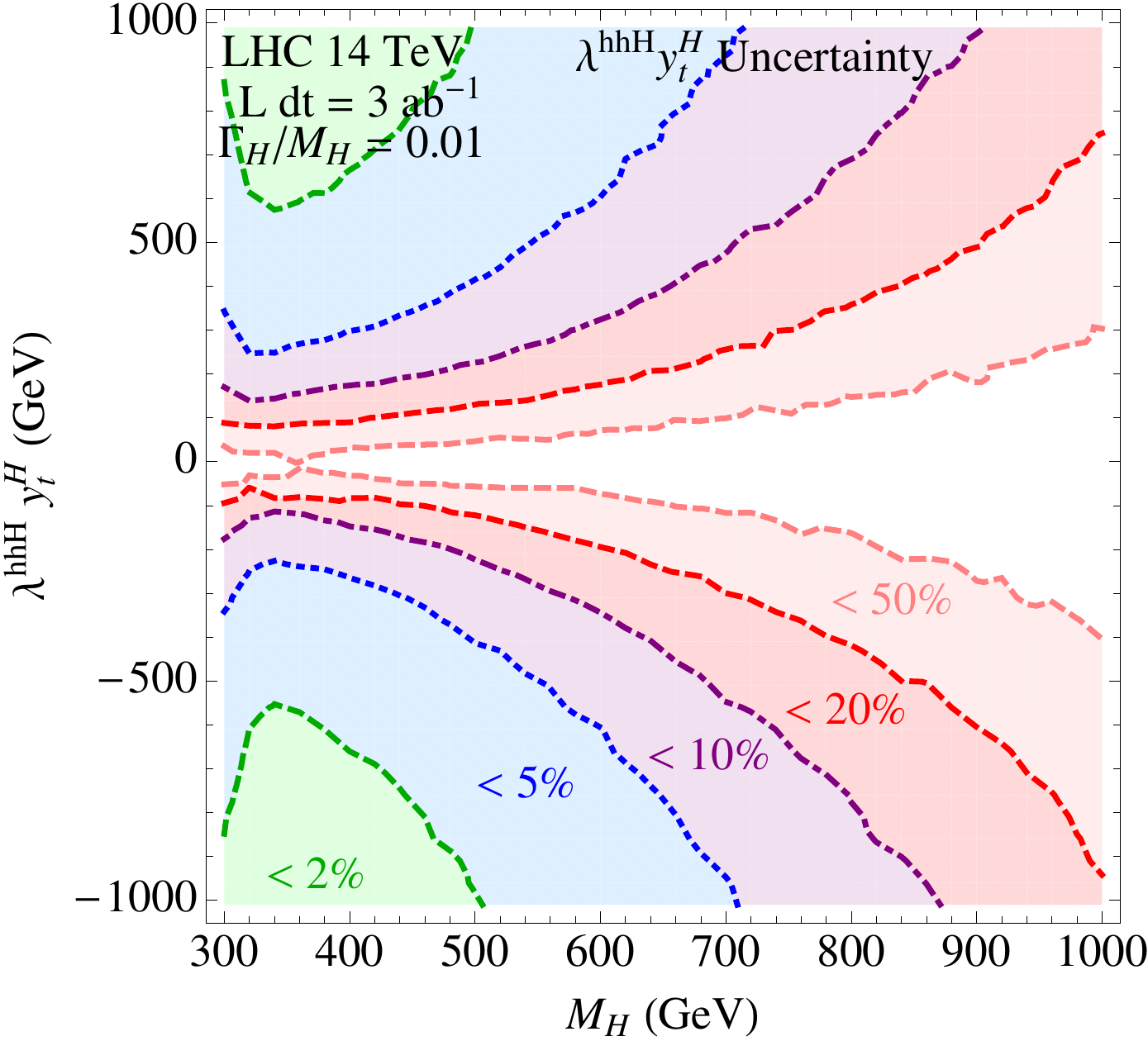}
\includegraphics[width=0.4\textwidth]{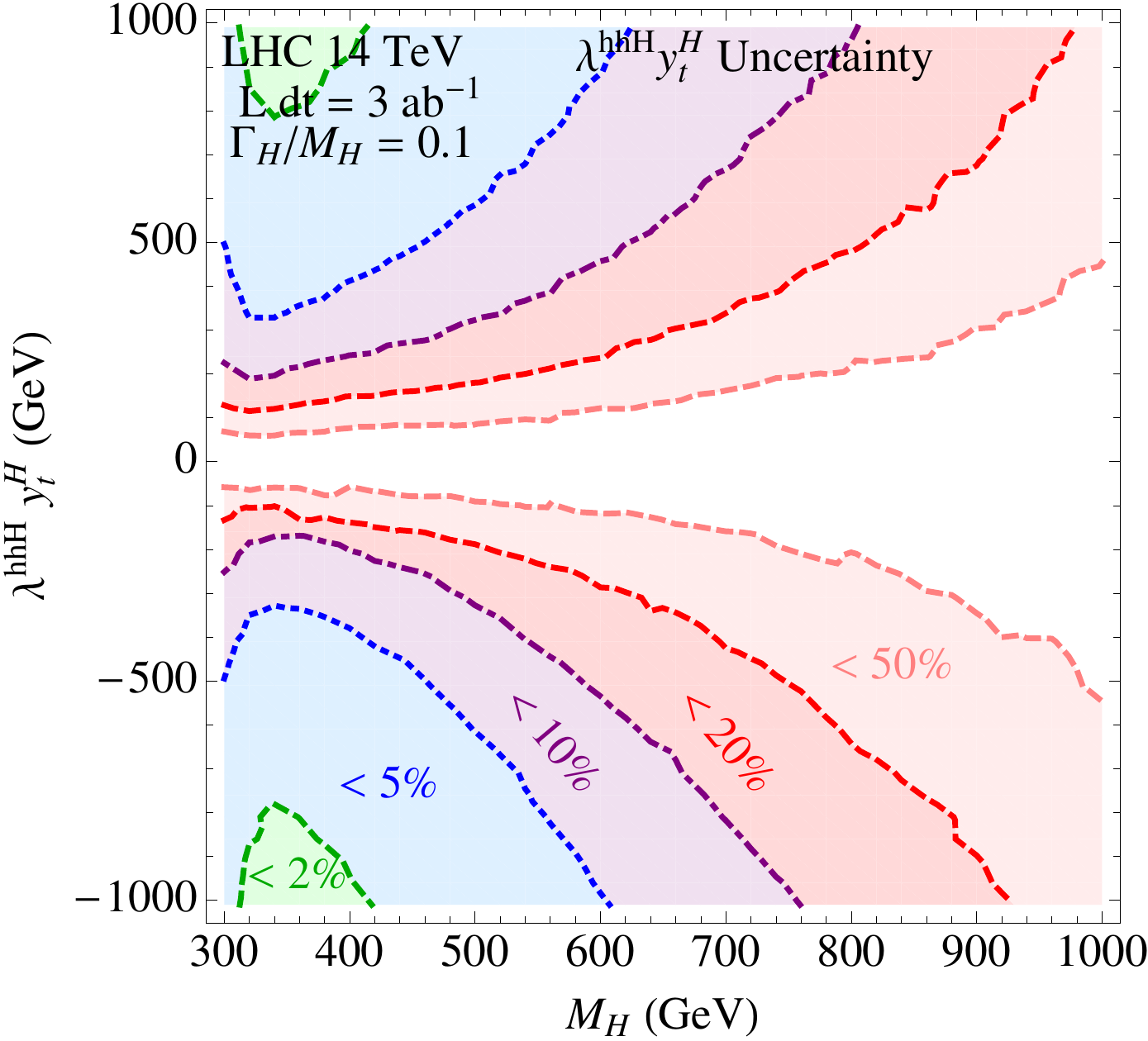}
\caption{Fit uncertainties at the $1\sigma$ level for $\lambda^{hhH} y_t^H$ for $\Gamma_H/M_H =$ 0.01 (top) and 0.1 (bottom) with $\int {\cal L} dt = 3$ ab$^{-1}$.}  
\label{fig:fitcoup}
\end{center}
\end{figure}

We perform a $\Delta \chi^2$ fit to approximately 1500 points in the $M_H-\lambda^{hhH} y_t^H$ plane for $\Gamma_H/M_H = 0.01$(0.1), varying $M_H$, $\Gamma_H$, $\lambda^{hhH} y_t^H,$ and $\lambda^{hhh}$ simultaneously and restricting the luminosity to ${\cal L} dt = 3$ ab$^{-1}$.  We find that $M_H$ can be measured to 0.5\% or better and the total width to 20\% or better over most of the parameter space.  For an intrinsically broader Higgs width with $\Gamma_H/M_H = 0.1$, the fit is degraded for the Higgs mass measurement, but the fit uncertainty in $\Gamma_H/M_H$ is not that different.

In Fig.~\ref{fig:fitcoup}, we show the $1\sigma$ uncertainty of $\lambda^{hhH} y_t^H$, which can be probed to a remarkably sensitive level over most of the parameter space and in some cases to less than 2\%.  This sensitivity degrades somewhat as the total width is increased, but the uncertainty remains low.  For a particular model of Yukawa couplings, the scalar coupling can be accurately deduced.   Even better, with an independent measurement of the $H$ coupling to the quark sector, the scalar coupling can be uniquely determined.

{\it Conclusions.} We studied resonant pair production of the 125 GeV Higgs $h$ for the final state in which one $h$ decays to $\gamma \gamma$ and the other to $ \bb$.  In new physics models with a heavy CP-even Higgs $H$,  the $s$-channel resonance amplitude of the $H$ interferes with the SM amplitudes.  We find that LHC14 sensitivity to the $H\to hh$ channel is possible if the magnitude of the product of the scalar and top quark Yukawa couplings $\lambda^{hhH} y_t^H$ is $O(100\text{ GeV})$ or larger (for $M_H\leq 1$ TeV) and its sign can be inferred from the $hh$ lineshape via interference effects.  $\lambda^{hhH} y_t^H$ can be measured to greater than $10-20\%$ accuracy over a broad range of $M_H\leq 1$ TeV.  Using the resonant enhancement of $hh$ production, the scalar coupling can be deduced or directly measured, yielding a deeper understanding of the scalar sector with potentially profound implications.

{\it Acknowledgements.}  V.~B., L.~L.~E, A.~P. and G.~S. are  supported by the U. S. Department of Energy under the contract DE-FG-02-95ER40896. 

\appendix

\end{document}